\documentclass[epj]{svjour}
\usepackage{graphics}
\providecommand{\Journal}[4] {#1 {\textbf {#2}}, #3 (#4)}
\providecommand{\CTP}{Commun. Theor. Phys. } %
\providecommand{\EPJA}{Eur. Phys. J. A } %
\providecommand{\MPLA}{Mod. Phys. Lett. A} %
\providecommand{\NPA}{Nucl. Phys. A } %
\providecommand{\NPB}{Nucl. Phys. B } %
\providecommand{\PL}{Phys. Lett. } %
\providecommand{\PLB}{Phys. Lett. B } %
\providecommand{\PR}{Phys. Rev. } %
\providecommand{\PRL}{Phys. Rev. Lett. } %
\providecommand{\PRC}{Phys. Rev. C } %
\providecommand{\PRD}{Phys. Rev. D } %
\providecommand{\RMP}{Rev. Mod. Phys. } %
\providecommand{\RMP}{Rev. Mod. Phys. } %
\providecommand{\ZPA}{Z. Phys. A } %
\begin{document}
\title{The 27-plet baryons with spin 3/2 under SU(3) symmetry}
\author{Qihua Zhou\inst{1} \and Bo-Qiang Ma\inst{2,1}
\thanks{\emph{email:} mabq@phy.pku.edu.cn}%
}                     
%
%
\institute{Department of Physics, Peking University, Beijing 100871,
China \and CCAST (World Laboratory), P.O.~Box 8730, Beijing 100080,
China}
\date{Received: date / Revised version: date}
%
\abstract{ We investigate the spin 3/2 baryons in the 27-plet based
on flavor SU(3) symmetry. For $J^p=3/2^+$, we find all the
candidates for non-exotic members. For $J^p=3/2^-$, we predict a new
non-exotic member $\Lambda(1780)$. Fitting the mass spectrum and
calculating the widths of the members show an approximate symmetry
of the 27-plet of SU(3). We find that the exotic members have
relatively large widths and the $\Xi(1950)$ has spin and parity
$J^p=3/2^-$. The possibility of assigning the non-exotic candidates
to an octet is also analyzed.
} 
\maketitle
\section{Introduction}
\label{intro} The SU(3) classification scheme proposed by Gell-Mann
and Ne'eman in 1961 has been proved quite successful and fruitful in
the investigation of hadron spectroscopy. In this classification
scheme, one can group the experimentally known strongly interacting
particles with the same quantum numbers of spin and parity into
various irreducible representations of the SU(3) group. There are
several SU(3) multiplets which have been well established by this
means, for example, $J^{P}$=$\frac{1}{2}^{+}$ octet and
$J^{P}$=$\frac{3}{2}^{+}$ decuplet baryons, which supplied clear and
unambiguous evidence for the SU(3) classification scheme. Higher
multiplets are also allowed in SU(3), such as $\bar{10}$, 27, 35,
etc. Because they contain the so-called exotic states beyond the
three quark $qqq$ content in the language of the conventional quark
model as Gell-Mann mentioned~\cite{Gell} and could hardly be found
in early year's experiments, the higher multiplet scheme received
little attention. From quantum chromodynamics (QCD), the underlying
theory of the strong interaction, the possibility for the existences
of exotic states can not be ruled out. Chiral soliton model ($\chi
SM$)~\cite{XSM} motivated investigations on the antidecuplet which
contains the exotic state $\Theta^+$ reported first by LEPS
Collaboration later~\cite{LEPS}. The $\Theta^+$ state has the
minimal quark content $uudd\bar{s}$~\cite{GnM} and hence is an
exotic pentaquark state with positive strangeness. Higher multiplet
27, which contains an isovector $\Theta$, also attracted some
attention. $\chi SM$ predicted a new isotriplet of $\Theta$
baryon~\cite{BFK,WnM1,WnM2}, with its mass being about 1.6 GeV and
width about 80 MeV. With the flux-tube quark model and the QCD sum
rules, Kanada-En'yo {\it et al.} predicted the $\Theta$ with
$I(J^p)=1(3/2^-)$ and mass $1.4\sim1.6$ GeV~\cite{KMN,NKMY}.
Noticeably, recently the STAR collaboration at RHIC presented data
indicating a small but significant $\Theta^{++}$ candidate with mass
about 1528 MeV~\cite{HZH}. Though it has more and more negative
reports against the existence of the $\Theta^+(1540)$ at present, it
is worthwhile to explore these new exotic particles.

In this paper, we examine possible non-exotic candidates of the
27-plet with spin $3/2$ in baryon particle listings from Particle
Data Group~\cite{PDG} by calculating their masses and partial decay
widths based on the approximate flavor SU(3) symmetry of the strong
interaction. Up to the present, seldom works about the exotics tried
to approach this issue using the most general and model-independent
method SU(3). In the quark model, these non-exotic candidates were
often assigned to the 56-plet of SU(6) with orbital excitation.
However, in Ref.~\cite{WnM1}, it could also get a rather good result
from chiral soliton model without demanding such analysis. It means
that we can try a more general analysis. Our motivation is an
attempt to verify whether the SU(3) symmetry, which has been greatly
successful in hadron physics~\cite{DeS,SGM}, can continue to play an
important role in the investigation of new particles. By this means,
Guzey and Polyakov have reviewed the spectrum of all baryons with
the mass less than approximately 2000-2200 MeV and catalogued them
into twenty-one SU(3) multiplet including 1, 8, 10 and
$\bar{10}$~\cite{GnP}. That work can be viewed as an update of
Ref.~\cite{SGM}. Likewise, the present study does not depend on any
specific model, neither does it introduce any presupposed adjustable
parameters. So, it is rather general and can be compared with
results of other works which are model-dependent.
\section{Mass spectrum and decay width}
\label{sec:1} 
\begin{figure}
\vspace{0.25cm}
\begin{center}
\resizebox{0.5\textwidth}{!}{
\includegraphics{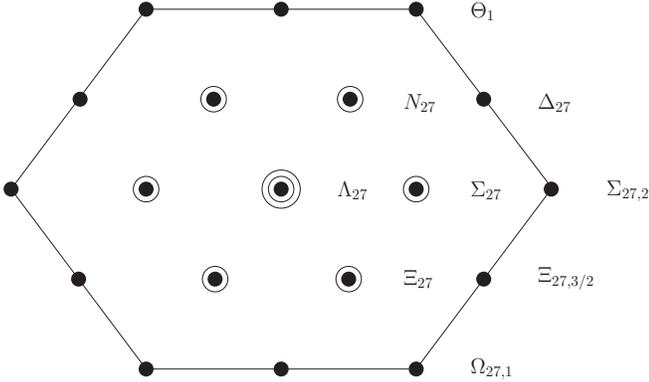}}
\end{center}
\vspace{0.25cm} \caption{Weight diagram for the 27-plet baryons.}
\label{fig1}
\end{figure}

The main tools of SU(3) systematization are the
well-known Gell-Mann--Okubo (GMO) mass formulae and the calculations
of two-body hadronic decays as showed in Refs.~\cite{DeS,SGM,GnP}.
We will follow this classical treatment process. Firstly, the masses
of baryons in the 27-plet can be obtained by using the GMO mass
formula:
\begin{equation}
   M=M_0+\alpha Y+\beta D^3_3,
\end{equation}
where $M_0$ is a common mass of a given multiplet and
$D^3_3=I(I+1)-Y^2/4-C/6$ with $C=2(p+q)+\frac{2}{3}(p^2+pq+q^2)$ for
the $(p,q)$ irreducible representation. $\alpha$ and $\beta$ are
mass constants that depend on the  representation which the baryon
belongs to. For 27-plet, $(p,q)$ is $(2,2)$, whose weight diagram
and the labels for the member states are shown in Fig.~\ref{fig1}.
Then we can get
\begin{eqnarray}
\Theta_1=M_0+2\alpha-\frac{5}{3}\beta,
&&\Delta_{27}=M_0+\alpha+\frac{5}{6}\beta, \nonumber\\
N_{27}=M_0+\alpha-\frac{13}{6}\beta,
&&\Sigma_{27}=M_0-\frac{2}{3}\beta, \nonumber\\
\Lambda_{27}=M_0-\frac{8}{3}\beta,
&&\Xi_{27}=M_0-\alpha-\frac{13}{6}\beta, \nonumber\\
\Sigma_{27,2}=M_0+\frac{10}{3}\beta,
&&\Xi_{27,3/2}=M_0-\alpha+\frac{5}{6}\beta, \nonumber\\
\Omega_{27,1}=M_0-2\alpha-\frac{5}{3}\beta.
\end{eqnarray}
From above, one can find some interesting relations, such as
the similar octet GMO relation \\
\begin{equation}
2N_{27}+\Xi_{27}=3\Lambda_{27}+\Sigma_{27}, \label{oct}
\end{equation}
and five independent equal-spacing rules \\
\begin{equation}
\begin{array}{c@{\,-\,}c@{\;=\;}c@{\,-\,}c}
\Theta_1 & \Delta_{27} & \Delta_{27}& \Sigma_{27,2}, \\
\Sigma_{27,2} & \Xi_{27,3/2} & \Xi_{27,3/2} & \Omega_{27,1}, \\
\Theta_1 & N_{27} & N_{27}& \Lambda_{27}, \\
\Lambda_{27} & \Xi_{27} & \Xi_{27}& \Omega_{27,1}, \\
N_{27} & \Xi_{27} & \Delta_{27} & \Xi_{27,3/2}\label{eqs}.
\end{array}
\end{equation}
In order to determine the mass spectrum, it just needs to know the
masses of three certain states in the 27-plet and requires that they
do not satisfy Eq.~(\ref{eqs}) at the same time. For the case of
$J^p=3/2^+$, we choose the following three well-established
resonances taken from PDG as inputs: $\Delta(1600)$, $N(1720)$ and
$\Lambda(1890)$. For the case of $J^p=3/2^-$, we choose
$\Delta(1940)$, $N(1720)$ and $\Sigma(1940)$. Other inputs are
possible, but give no more candidates in PDG than themselves. The
best mass fitting results are shown in Table~\ref{tab1}. Note that
we now have two sets of 27-plet baryons. For the set of $J^p=3/2^+$,
all the non-exotic states have their candidates. For the set of
$J^p=3/2^-$, we get a new state $\Lambda_{27}$ with mass 1780 MeV.
The existence of new $\Lambda$ hyperon with $J^p=3/2^-$ was
predicted in specific constituent quark models with various
assumptions about the quark dynamics~\cite{GPVW,LMP}.
Ref.~\cite{GPVW} gave the mass about 1780 MeV and model A in
Ref.~\cite{LMP} gave the mass of 1775 MeV. Here, we directly get it
only by completing the SU(3) picture of the 27-plet baryons. This
result comes from model-independent analysis, which just shows the
merit of SU(3). In both sets, all exotic states have no candidates
at present.

Next, we will calculate the two-body partial widths of the 27-plet
baryons decaying to the octet baryons and the pseudoscalar mesons to
verify this assignment. The SU(3) invariant 27-8-8 interaction
Lagrangian can be obtained by constructing the SU(3) singlet like
this form
\begin{equation}
L=g_{27}\bar{T}_{ij}^{kl}B^i_kM^j_l\label{lag},
\end{equation}
where $T_{ij}^{kl}$ is an irreducible tensor notation to represent
the 27-plet baryons, and $B^i_k$ and $M^j_l$ denote the baryon octet
and pseudoscalar meson octet respectively. The full expression
written in terms of the physical states has been deduced
in~\cite{OnH}. For the concrete decay process of a 27-plet baryon
$B^\prime$ with spin $3/2$ to an octet baryon $B$ with spin $1/2$
and a pseudoscalar meson $M$
\begin{equation}
B^\prime\rightarrow B+M,
\end{equation}
the calculation can be performed in the framework of
Rarita-Schwinger formalism. The parity-conserving interaction
Lagrangian are~\cite{Rush,Carr}
\begin{eqnarray}
L_+ &=& \frac{g_{B^\prime
BM}}{m_\pi}\bar{\psi}\Psi^\mu\partial_\mu\phi, \nonumber \\
L_- &=& i\frac{g_{B^\prime BM}}{m_\pi}\bar{\psi}\gamma_5\Psi^\mu
\partial_\mu\phi,
\end{eqnarray}
for $B^\prime$ with positive or negative parity respectively, where
$\psi$ is the spin $1/2$ field and $\Psi^\mu$ is the spin $3/2$
field. The pseudoscalar meson field is $\phi$ and the factor
$1/{m_\pi}$ is introduced to make the coupling constant $g_{B^\prime
BM}$ relative to the universal coupling constant $g_{27}$ in
Eq.~(\ref{lag}) dimensionless. We can get the coupling constant
$g_{B^\prime BM}$ by directly computing the Clebsch-Gordan
coefficient among the SU(3) irreducible representations of
$B^\prime,B,M$. Accordingly, the decay widths are written as
\begin{eqnarray}
    \Gamma_+(B^\prime\rightarrow B M) &=& \frac{g^2_{B^\prime B
m}}{12\pi
m_\pi^2}p^3\frac{\left[(m_{B^\prime}+m_B)^2-m^2\right]}{m_{B^\prime}^2},
\nonumber \\
    \Gamma_-(B^\prime\rightarrow B M) &=& \frac{g^2_{B^\prime B
m}}{3\pi m_\pi^2}p^5\frac{1}{\left[(m_{B^\prime}+m_B)^2-m^2\right]},
\end{eqnarray}\label{width}
where $p$ is the c.m. momentum value of the final meson. In terms of
the baryons masses $m_{B^\prime}$, $m_B$ and the meson mass $m$, we
have
\begin{equation}
p=\frac{\sqrt{\left[(m_{B^\prime}+m_B)^2-m^2\right]\left[(m_{B^\prime}-m_B)^2-m^2\right]}}{2m_{B^\prime}}.
\end{equation}
After some trivial calculations, we get the results expressed by the
universal coupling constant $g_{27}^2$ and list them particularly in
Table~\ref{tab1}. To examine whether the SU(3) symmetry can hold, we
need to compare all the ratios of two certain partial decay widths
in Table~\ref{tab1} with the data from experiment. Here, we choose
the minimum experimental value of $\Gamma(N_{27}\rightarrow N\pi)$
as input just to show the validity. Similarly, other choices can
easily be verified. We also list the results in Table~\ref{tab1}.
\begin{table*}
\centering
\caption{\label{tab1}The masses and widths of baryons in
the 27-plet (in unit of MeV)}
\begin{tabular}{cccccccc}
\hline
& Candidate & Width in PDG  & Decay mode & Branching ratio &
$\Gamma_i(exp)$  & $\Gamma_i(th)$  & $\Gamma_i(th)$
[$g^2_{27}$] \\
\hline
$J^p=3/2^+$ \\
\hline
$\Delta_{27}(1600)$ & $\Delta(1600)$ & 250-450 & $N\pi$ & 10\%-25\% & 25-112.5 & 129 & 3579.3 \\
$N_{27}(1720)$ & $N(1720)$ & 100-200 & $N\pi$ & 10\%-20\% & 10-40 & 10(input) & 277.5 \\
& & & $N\eta$ & $(4.0\pm1.0)\%$ & 3-10 & 30.9 & 857.1 \\
& & & $\Lambda{K}$ & 1\%-15\% & 1-30 & 10.5 & 291.7 \\
& & & $\Sigma{K}$ & & & 0.2 & 5.1 \\
$\Sigma_{27}(1810)$ & $\Sigma(1840)$ & $120\pm10$ & $N\bar{K}$ & $0.37\pm0.13$ & 26.4-65 & 46.5 & 1290.4 \\
$\Lambda_{27}(1890)$ & $\Lambda(1890)$ & 60-200 & $N\bar{K}$ & 20\%-35\% & 12-70 & 14 & 388.2 \\
& & & $\Sigma\pi$ & 3\%-10\% & 1.8-20 & 2.3 & 63.8 \\
$\Xi_{27}(2020)$  & $\Xi(2030)$ & $20^{+15}_{-5}$ & $\Lambda\bar{K}$ & $\sim$20\% & 3-7 & 85.9 & 2384.8 \\
& & & $\Sigma\bar{K}$ & $\sim$80\% & 12-28 & 7.3 & 202.9 \\
$\Theta_1(1550)$ & $?$ & ? & $NK$ & ? & ? & 33.2 & 920.1 \\
$\Sigma_{27,2}(1650)$ & $?$ & ? & $\Sigma\pi$ & ? & ? & 164.4 & 4562.1 \\
$\Xi_{27,3/2}(1900)$ & $?$ & ? & $\Xi\pi$ & ? & ? & 125.8 & 3491.2 \\
& & & $\Sigma\bar{K}$ & ? & ? & 3.4 & 95.5 \\
$\Omega_{27,1}(2150)$ & $?$ & ?& $\Xi\bar{K}$ & ? & ? & 232.9 & 6461.8 \\
\hline
$J^p=3/2^-$ \\
\hline
$\Delta_{27}(1940)$ & $\Delta(1940)$ & $460\pm320$ & $N\pi$ & $0.18\pm0.12$ & 8.4-234 & 136.9 & 1155.2 \\
& & & $\Sigma{K}$ & & & 9.7 & 81.7 \\
$N_{27}(1700)$ & $N(1700)$ & 50-150 & $N\pi$ & 5\%-15\% & 2.5-22.5 & 2.5(input) & 21.1 \\
& & & $N\eta$ & $(0.0\pm1.0)\%$ & 0-1.5 & 3.7 & 31.3 \\
& & & $\Lambda{K}$ & $<$3\% & $<$4.5 & 0.3 & 2.7 \\
& & & $\Sigma{K}$ & & & 0.0002 & 0.002 \\
$\Sigma_{27}(1940)$ & $\Sigma(1940)$ & 150-300 & $N\bar{K}$ & $<$20\% & $<$60 & 20.5 & 172.6 \\
$\Lambda_{27}(1780)$ & $?$ & ? & $N\bar{K}$ & ? & ? & 0.3 & 16.2 \\
& & & $\Sigma\pi$ & ? & ? & 0.18 & 1.5 \\
$\Xi_{27}(1940)$  & $\Xi(1950)$ & $60\pm20$ & $\Lambda\bar{K}$ & seen & & 9.9 & 83.8 \\
& & & $\Sigma\bar{K}$ & possibly seen & & 0.5 & 4.5 \\
& & & $\Xi\pi$ & seen & & 0.9 & 7.7 \\
$\Theta_1(1620)$ & $?$ & ? & $NK$ & ? & ? & 48.9 & 412.3 \\
$\Sigma_{27,2}(2260)$ & $?$ & ? & $\Sigma\pi$ & ? & ? & 400 & 3375.9 \\
$\Xi_{27,3/2}(2180)$ & $?$ & ? & $\Xi\pi$ & ? & ? & 63.8 & 538.5 \\
& & & $\Sigma\bar{K}$ & ? & ? & 56.4 & 475.6 \\
$\Omega_{27,1}(2100)$ & $?$ & ?& $\Xi\bar{K}$ & ? & ? & 19 & 160 \\
\hline
\end{tabular}
\end{table*}

\section{Discussion}
Up to now, we only considered the pure 27-plet assignment. Actually,
the 27-plet can mix with other representations, such as the octet or
the decuplet with spin $3/2$ equally. Because the well-established
$J^p=3/2^+$ decuplet works well for the mass spectrum, it will imply
that the mixing with the 27-plet is small. Possibly, mixing can take
place between the set of the 27-plet with $J^p=3/2^-$ and the
potentially octet with the same quantum numbers in the available
particle listings~\cite{HnH}. But from Ref.~\cite{GnP}, the possible
pure octet assignment of $N(1520)$, $\Lambda(1690)$, $\Sigma(1670)$
and $\Xi(1820)$ seems to work well too. So, we do not deeply treat
the mixing here.

Because of the similar octet GMO relation Eq.~(\ref{oct}), it seems
that $N_{27},\Sigma_{27},\Lambda_{27}$ and $\Xi_{27}$ can be
assigned to a pure octet. We need to calculate their partial decay
widths to examine this possibility. Again, we construct the SU(3)
invariant 8-8-8 interaction Lagrangian by two possible couplings,
namely, the well-known $f$- and $d$-type interactions. We write the
interaction Lagrangian as
\begin{equation}
L=g_8(d+f)\bar{P}^l_iB^i_kM^k_l+g_8(d-f)\bar{P}^l_iB^k_lM^i_k,
\end{equation}
where $\bar{P}^l_i$ represents the octet consists of
$N_{27},\Sigma_{27},\Lambda_{27}$ and $\Xi_{27}$, and $B^i_k$ and
$M^k_l$ denote the baryon octet and the pseudoscalar meson octet
respectively. After an analogous process as the foregoing, we list
the results expressed by the universal coupling constant $g_8$ and
two parameters $f,d$ in Table~\ref{tab2}. The appropriate fitting
results of these partial decay widths are also presented. Compared
with the results from the 27-plet, the picture of the octet seems to
be able to give right relative magnitudes of the partial decay
widths of a certain baryon. For example, $\Gamma(N_{27}\rightarrow
N\pi)$ should be broader than $\Gamma(N_{27}\rightarrow N\eta)$
according to the experimental data while the picture of 27-plet
gives the reverse results both in SU(3) and $\chi SM$~\cite{WnM1}.
But, there are still some partial decay widths such as
$\Gamma(\Sigma_{27}\rightarrow N\bar{K})$ which are depressed too
low as can be seen from Table~\ref{tab2}. Simultaneously, note that
the analysis~\cite{LMP} predicts that the $\Lambda$ couples very
weakly to the $N\bar{K}$ state. The calculation from 27-plet
supports this to be more reasonable than that from the octet.
Although no appropriate $f/d$ ratio can be found to be compatible
with all experimental data, the picture of octet still can not be
completely excluded especially for the case of $J^p=3/2^-$ because
of the large widths and the imprecise branch ratios of its members.
It will need more exact experimental data to examine this
possibility. Of course, the picture of 27-plet is more attractive,
as it provides also the connection of $\Delta(1600)$ and
$\Delta(1940)$ with other 27-plet members with $J^{P}=3/2^{+}$ and
$J^{P}=3/2^{-}$ respectively, especially those in the center of
weight diagram of 27-plet. So it contains more information than the
picture of octet. $\Theta^{++}$ can also be contained in higher
multiplet such as 35-plet. However, we know that for 35-plet, any
member of which can not decay to a octet baryon and a pseudoscalar
meson because of the confinement coming from the group theory.
Actually, many particles in PDG as potential candidates have such
large decay branch. Therefore we think that the existence of
multiplet higher than 27-plet is unlikely possible under the SU(3).
For the possible $\Theta^{++}$ referred in Ref.~\cite{KMN,NKMY,HZH},
the 27-plet is very hopeful of containing it.
\begin{table*}
\centering
\caption{\label{tab2}The widths of baryons in the octet
(in unit of MeV)}
\begin{tabular}{cccccccc}
\hline
& Candidate & Width in PDG  & Decay mode & Branching ratio & $\Gamma_i(exp)$  & $\Gamma_i(th)$  & $\Gamma_i(th)$ [$g^2_8$] \\
\hline
$J^p=3/2^+$ \\
\hline
$N_{27}(1720)$ & $N(1720)$ & 100-200 & $N\pi$ & 10\%-20\% & 10-40 & 27.4 & $2081.3(d+f)^2$ \\
& & & $N\eta$ & $(4.0\pm1.0)\%$ & 3-10 & 1.8 & $79.4(d-3f)^2$ \\
& & & $\Lambda{K}$ & 1\%-15\% & 1-30 & 1.7 & $27(d+3f)^2$ \\
& & & $\Sigma{K}$ & & & 0.01 & $38.3(d-f)^2$ \\
$\Sigma_{27}(1810)$ & $\Sigma(1840)$ & $120\pm10$ & $N\bar{K}$ & $0.37\pm0.13$ & 26.4-65 & 0.4 & $1613(d-f)^2$ \\
$\Lambda_{27}(1890)$ & $\Lambda(1890)$ & 60-200 & $N\bar{K}$ & 20\%-35\% & 12-70 & 13.2 & $215.7(d+3f)^2$ \\
& & & $\Sigma\pi$ & 3\%-10\% & 1.8-20 & 3 & $1276d^2$ \\
$\Xi_{27}(2020)$  & $\Xi(2030)$ & $20^{+15}_{-5}$ & $\Lambda\bar{K}$ & $\sim$20\% & 3-7 & 5(input) & $220.7(d-3f)^2$ \\
& & & $\Sigma\bar{K}$ & $\sim$80\% & 12-28 & 20(input) & $1521.8(d+f)^2$ \\
\hline
$J^p=3/2^-$ \\
\hline
$N_{27}(1700)$ & $N(1700)$ & 50-150 & $N\pi$ & 5\%-15\% & 2.5-22.5 & 10(input) & $158.3(d+f)^2$ \\
& & & $N\eta$ & $(0.0\pm1.0)\%$ & 0-1.5 & 1(input) & $2.9(d-3f)^2$ \\
& & & $\Lambda{K}$ & $<$3\% & $<$4.5 & 0.1 & $0.3(d+3f)^2$ \\
& & & $\Sigma{K}$ & & & 0.001 & $0.02(d-f)^2$ \\
$\Sigma_{27}(1940)$ & $\Sigma(1940)$ & 150-300 & $N\bar{K}$ & $<$20\% & $<$60 & 6.1 & $215.8(d-f)^2$ \\
$\Lambda_{27}(1780)$ & $?$ & ? & $N\bar{K}$ & ? & ? & 4 & $9(d+3f)^2$ \\
& & & $\Sigma\pi$ & ? & ? & 0.1 & $30d^2$ \\
$\Xi_{27}(1940)$  & $\Xi(1950)$ & $60\pm20$ & $\Lambda\bar{K}$ & seen & & 2.7 & $7.8(d-3f)^2$ \\
& & & $\Sigma\bar{K}$ & possibly seen & & 15.2 & $33.8(d+f)^2$ \\
& & & $\Xi\pi$ & seen & & 1.6 & $57.8(d-f)^2$ \\
\hline
\end{tabular}
\end{table*}

%
%
%

\section{Summary}
In summary, we use the flavor SU(3) symmetry to examine the possible
candidates of 27-plet with spin 3/2. By calculating the partial
decay widths of the candidates, the approximate symmetry of the
27-plet of SU(3) can be seen.
For $J^p=3/2^-$ multiplet, we predict a new missing baryon
$\Lambda(1780)$, no matter in the picture of 27-plet or octet. The
picture of 27-plet provides also the connection of  $\Delta(1600)$
and $\Delta(1940)$ with other members of $J^{P}=3/2^{+}$ and
$J^{P}=3/2^{-}$ baryons. Compared with the results from $\chi SM$,
the non-exotic candidate $\Xi(1950)$ has $J^p=3/2^-$, which was
predicted to be $J^p=3/2^+$ in Ref.~\cite{WnM1}. In both cases, one
can find that the exotic members have relatively larger widths than
those of non-exotic members, which makes them more difficult to be
detected experimentally. The results obtained here are model
independent, and would be useful for the future study of new baryons
by combining with other dynamical approaches.


We are grateful to Bin Wu for useful discussions. This work is
partially supported by National Natural Science Foundation of China
(Nos.~10421503, 10575003, 10528510), by the Key Grant Project of
Chinese Ministry of Education (No.~305001), by the Research Fund for
the Doctoral Program of Higher Education (China).

%

\begin{thebibliography}{}
%
%

\bibitem{Gell}
M.~Gell-Mann, \Journal{\PL} {8}{214}{1964}.

\bibitem{XSM}
A.V.~Manohar, \Journal{\NPB} {248}{19}{1984}; M.~Chemtob,
\Journal{\NPB} {256}{600} {1985}; H.~Walliser, \Journal{\NPA}
{548}{649} {1992}; M.-L.~Yan and X.-H.~Meng, \Journal{\CTP}
{24}{435} {1995}; D.~Diakonov, V.~Petrov and M.~Polyakov,
\Journal{\ZPA} {359}{305}{1997}.

\bibitem{LEPS}
LEPS, T.~Nakano, {\it et al.}, \Journal{\PRL} {91}{012002}{2003}.

\bibitem{GnM}
H.~Gao and B.-Q.~Ma, \Journal{\MPLA} {14}{2313}{1999}.

\bibitem{BFK}
D.~Borisyuk, M.~Faber and A.~Kobushkin, hep-ph/0307370.

\bibitem{WnM1}
B.~Wu and B.-Q.~Ma, \Journal{\PRD}{69}{077501}{2004}.

\bibitem{WnM2}
B.~Wu and B.-Q.~Ma, \Journal{\PLB}{586}{62}{2004}.

\bibitem{KMN}
Y.~Kanada-En'yo, O.~Morimatsu and T.Nishikawa, \Journal{\PRC}
{71}{045202}{2005}.

\bibitem{NKMY}
T.~Nishikawa, Y.~Kanada-En'Yo, O.~Morimatsu and Y.~Kondo,
\Journal{\PRD} {71}{076004}{2005}.

\bibitem{HZH}
H.Z.~Huang, nucl-ex/0509037.

\bibitem{PDG}
Particle Data Group, S.~Eidelman, {\it et al.},
\Journal{\PLB}{592}{1}{2004}.

\bibitem{DeS}
J.J.~de~Swart, \Journal{\RMP} {35}{916}{1963}.

\bibitem{SGM}
N.P.~Samios, M.~Goldberg and B.T.~Meadows, \Journal{\RMP}
{46}{49}{1974}.

\bibitem{GnP}
V.~Guzey and M.V.~Polyakov, hep-ph/0512355.

\bibitem{GPVW}
L.Y.~Glozman, W.~Plessas, K.~Varga and R.F.~Wagenbrunn,
\Journal{\PRD} {58}{094030}{1998}.

\bibitem{LMP}
U.~Loring, B.C.~Metsch and H.R.~Petry, \Journal{\EPJA}
{10}{447}{2001}.

\bibitem{OnH}
Y.~Oh and H.~Kim, \Journal{\PRD} {70}{094022}{2004}.

\bibitem{Rush}
J.G.~Rushbrooke, \Journal{\PR} {143}{1345}{1966}.

\bibitem{Carr}
P.~Carruthers, \Journal{\PR} {152}{1345}{1966}.

\bibitem{HnH}
T.~Hyodo and A.~Hosaka, \Journal{\PRD} {71}{054017}{2005}.

\end{thebibliography}
%

\end{document}